\documentclass[aps,prl,reprint,superscriptaddress,showpacs]{revtex4-1}

\usepackage{graphicx,amsmath,amssymb,ulem}
\usepackage{verbatim}
\usepackage{ragged2e}
\usepackage{color}

\begin{document}

\title{Ring-shaped spectra of parametric down-conversion \\and entangled photons that never meet}

\author{Kirill Yu.~Spasibko}
\affiliation{Max-Planck Institute for the Science of Light, \\  Guenther-Scharowsky-Str. 1 / Bau 24, Erlangen  D-91058, Germany}
\affiliation{University of Erlangen-N\"urnberg, Staudtstrasse 7/B2, 91058 Erlangen, Germany}
\affiliation{Physics Department, Moscow State University, Leninskiye Gory 1-2, Moscow 119991, Russia}
\author{Denis~A.~Kopylov}
\affiliation{Physics Department, Moscow State University, Leninskiye Gory 1-2, Moscow 119991, Russia}
\author{Tatiana V.~Murzina}
\affiliation{Physics Department, Moscow State University, Leninskiye Gory 1-2, Moscow 119991, Russia}
\author{Gerd Leuchs}
\affiliation{Max-Planck Institute for the Science of Light, \\  Guenther-Scharowsky-Str. 1 / Bau 24, Erlangen  D-91058, Germany}
\affiliation{University of Erlangen-N\"urnberg, Staudtstrasse 7/B2, 91058 Erlangen, Germany}
\affiliation{Physics Department, Moscow State University, Leninskiye Gory 1-2, Moscow 119991, Russia}
\author{Maria V.~Chekhova}
\affiliation{Max-Planck Institute for the Science of Light, \\  Guenther-Scharowsky-Str. 1 / Bau 24, Erlangen  D-91058, Germany}
\affiliation{University of Erlangen-N\"urnberg, Staudtstrasse 7/B2, 91058 Erlangen, Germany}
\affiliation{Physics Department, Moscow State University, Leninskiye Gory 1-2, Moscow 119991, Russia}

\begin{abstract}
We report on the observation of an unusual type of parametric down-conversion. In the regime where collinear degenerate emission is in the anomalous range of group-velocity dispersion, its spectrum is restricted in both angle and wavelength. Detuning from exact collinear-degenerate phasematching leads to a ring shape of the wavelength-angular spectrum, suggesting a new type of spatiotemporal coherence and entanglement of photon pairs. By imposing a phase varying in a specific way in both angle and wavelength, one can obtain an interesting state of an entangled photon pair, with the two photons being never at the same point at the same time.
\end{abstract}

\maketitle

Parametric down-conversion (PDC) is the main source of entangled photon pairs and twin beams. It is also used in optical parametric oscillators as a source of tunable radiation. Its importance for quantum and nonlinear optics can hardly be over-estimated.

\begin{figure}[htbp]
\centering
\includegraphics[width=\linewidth]{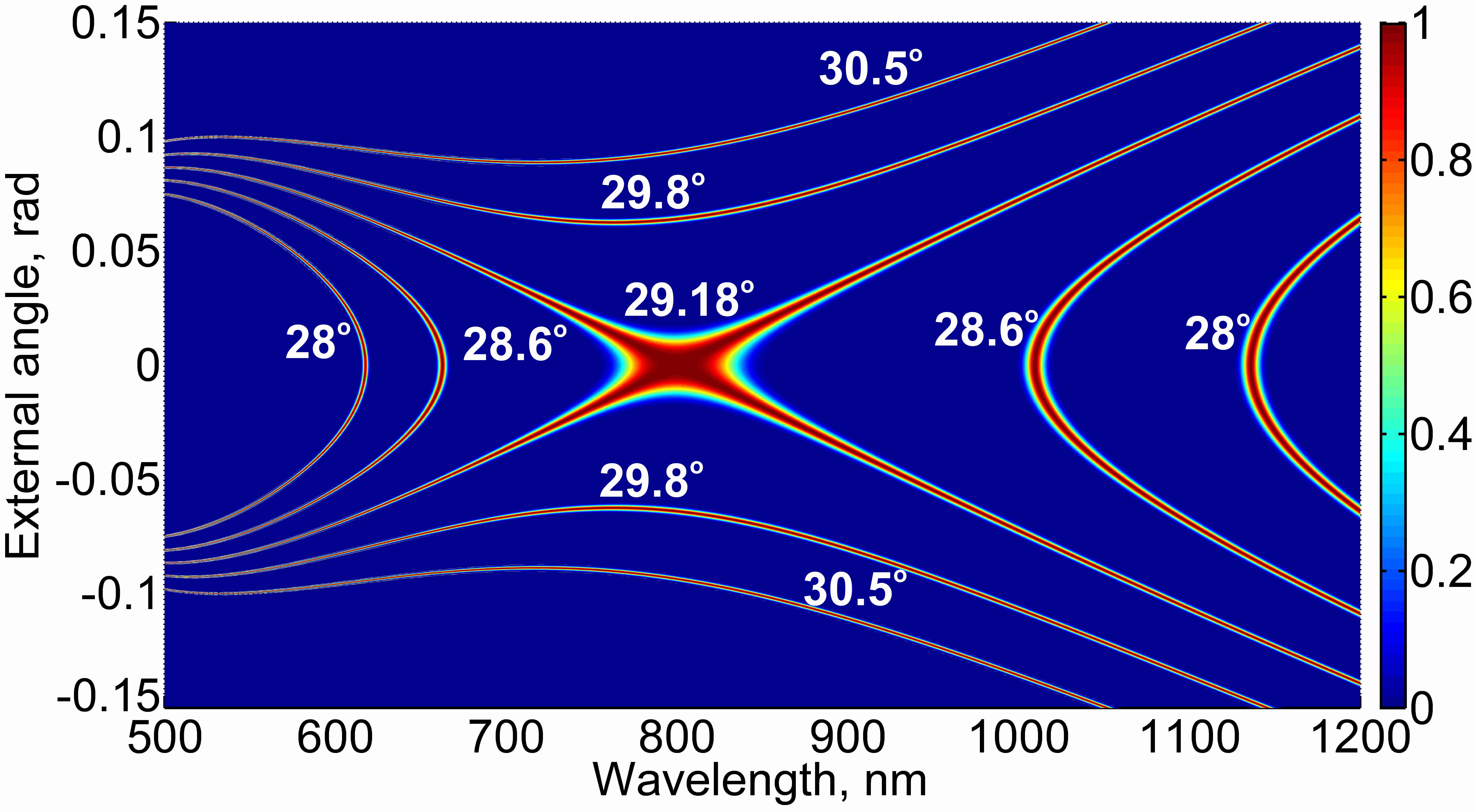}
\caption{Wavelength-angular spectra of type-I PDC in the case of 10 mm BBO crystal pumped at 400 nm for different angles between the pump wavevector and the optic axis (shown in white)}.
\label{fig:tuning}
\end{figure}

PDC has a very specific shape of the wavelength-angular spectra. 
%As known from the early time of parametric amplifiers, there is a relatively strict relation between the radiation angle of emission and its wavelength. 
So far, two types of such spectra have been observed. One, corresponding to type-I PDC, has a typical shape of a `cross' around the degenerate collinear point and becomes similar to two hyperbolas as the crystal orientation changes (Fig.~\ref{fig:tuning}). This `X-shape' of the spectrum is known since the 1970-s and has been many times described in the literature since then ~\cite{big,Piskarskas,X-ent2}. Importantly, the spectrum is practically unbounded, both in the wavelength and in the angle. In the other well-known case, the one of type-II PDC~\cite{type-II}, the spectra are also almost unbounded but do not have the typical `X' shape.

It has been shown theoretically~\cite{O-theory,O-theory2} that another type of wavelength-angular spectra exists for type-I PDC. Namely, if the degenerate wavelength coincides with the zero dispersion wavelength (ZDW), the frequency spectrum of collinear type-I PDC is extremely broadband, due to the phase mismatch depending only on higher powers of the frequency mismatch. This fact is used in works on parametric generation and amplification~\cite{Kuo2006,Tiihonen2006,Lim2007}. Even more interesting is the fact that, for PDC both at ZDW and beyond it (in the anomalous group-velocity dispersion (GVD) range), the wavelength-angular spectrum is bounded. This happens because in the anomalous GVD range, frequency detuning from the degenerate collinear PDC cannot be compensated by the non-collinear emission, as in the normal GVD range. As a result, collinear frequency-degenerate type-I PDC has a 'spot-like' wavelength-angular spectrum~\cite{ODonnell2007} and the non-degenerate/noncollinear phasematching leads to a ring-shaped spectrum~\cite{O-theory,O-theory2}.

Below, we demonstrate the first experimental observation of the ring-type of wavelength-angular spectrum, discuss its most interesting features and suggest some important applications.

\begin{figure}[htbp]
\centering
\includegraphics[width=0.9\linewidth]{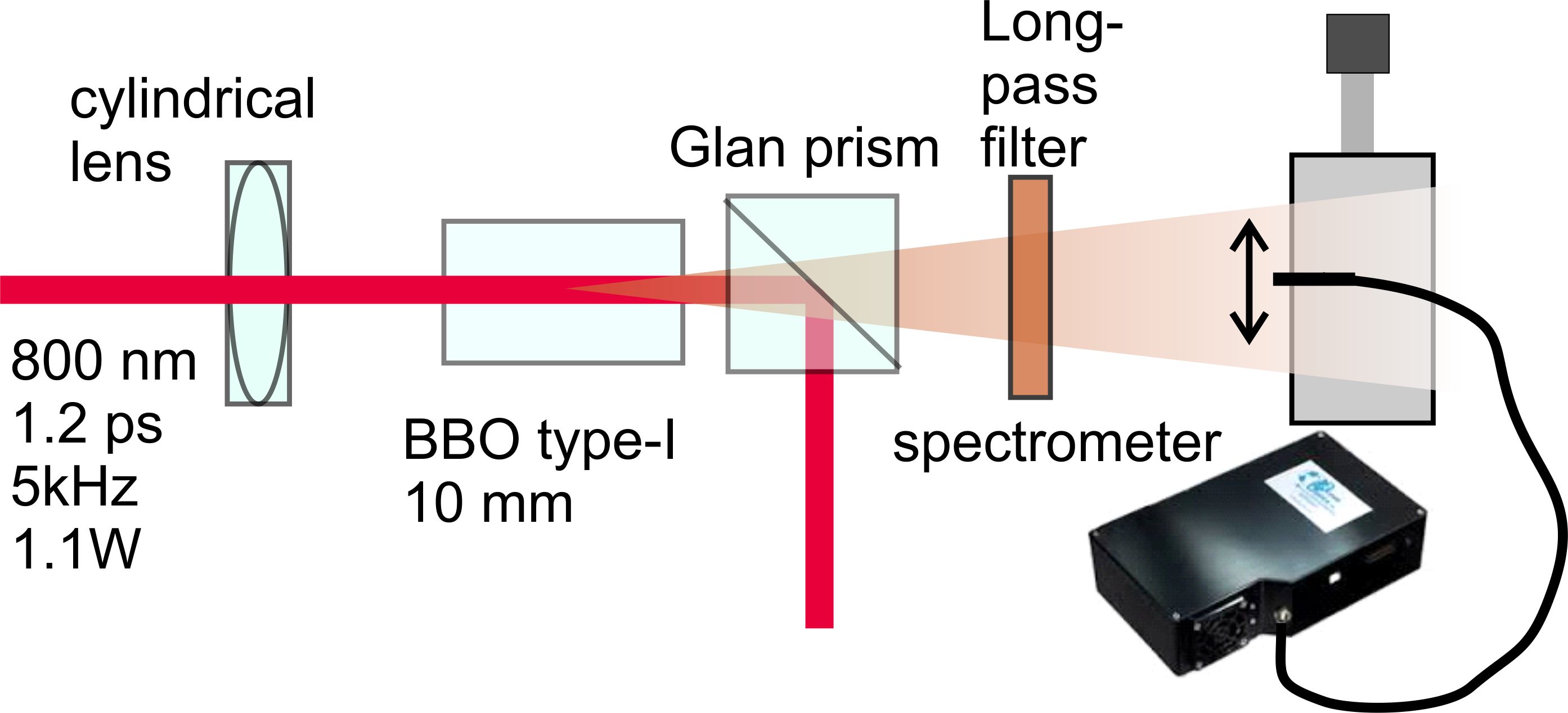}
\caption{The experimental setup for measuring wavelength-angular spectra of PDC.}
\label{fig:setup}
\end{figure}

\begin{figure}[htbp]
\centering
\includegraphics[width=0.9\linewidth]{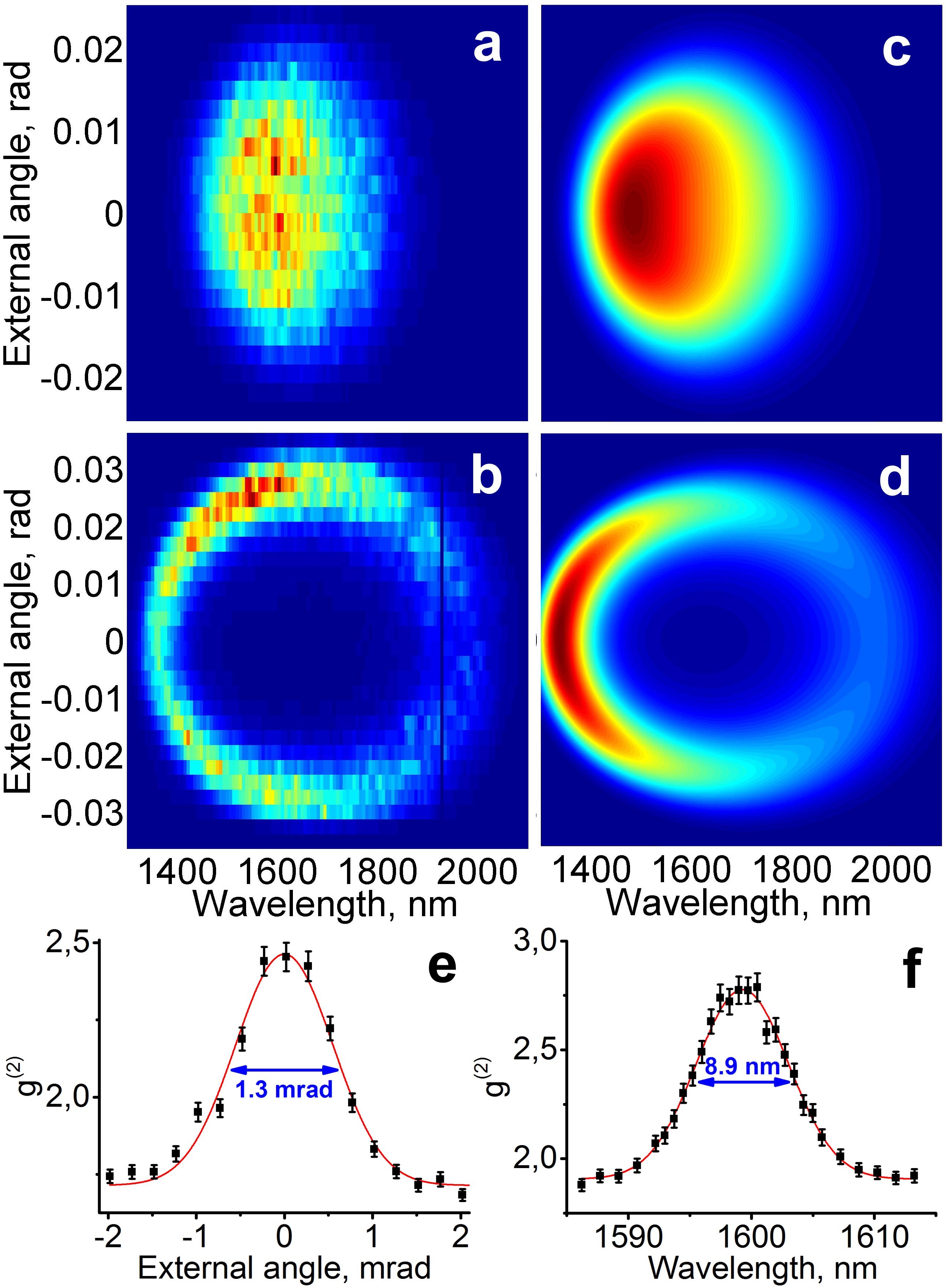}
\caption{Top: experimental (a,b) and theoretical (c,d) wavelength-angular spectra of PDC around $1.6\mu$m for the crystal orientations $19.87^\circ$ (a,c) and $19.98^\circ$ (b,d). The color code is the same as in Fig.~\ref{fig:tuning}. Bottom: the second-order normalized correlation function versus the angle (e) and the wavelength (f) for the same crystal orientation as in panels a,c.}
\label{fig:spectra_PDC}
\end{figure}

As a pump, we use the radiation of an amplified Ti-sapphire laser with the central wavelength $800$ nm, pulse duration $1.2$ ps, and repetition rate $5$ kHz. For the simplicity of detection, we work in the high-gain regime of PDC achieved by using a mean power of $1.1$ W. PDC is generated in a $10$-mm BBO crystal cut at $19.9^\circ$ to the optic axis (Fig.~\ref{fig:setup}). To avoid spatial walk-off effects, the pump radiation is focused into the crystal by means of a cylindrical lens with the focal length $700$ mm, so that no focusing takes place in the plane of the optic axis. The pump radiation is cut off by means of a Glan prism and a long-pass filter (FEL1200), and the wavelength-angular spectra of PDC are measured with the help of an IR fibre spectrometer (Ocean Optics NIRQuest256), with the fibre tip ($600\, \mu$m) placed at a distance of $40$ cm from the crystal. By displacing the fibre tip and each time recording the wavelength spectrum, we obtain wavelength spectra at different angles. The angular scanning is done in the horizontal direction, in which the pump beam is not focused. By combining the wavelength spectra at different angles into a two-dimensional plot, we get the two-dimensional spectra shown in Fig.~\ref{fig:spectra_PDC}a,b. The first one (a) was obtained for the crystal orientation $19.87^\circ$ and the second one (b), for $19.98^\circ$. The corresponding theoretical spectra are presented in  Fig.~\ref{fig:spectra_PDC}c,d. In the calculation, the high parametric gain was taken into account using the approach of Ref. \cite{Brambilla2004}.

The spectra are indeed restricted in both wavelength and angle. At exact collinear frequency-degenerate phase matching, the spectrum has the shape of a spot (Fig.~\ref{fig:spectra_PDC}a,c) while with the crystal detuned from this orientation, the shape resembles a ring (Fig.~\ref{fig:spectra_PDC}b,d). The intensity decrease towards long wavelengths is due to the fact that for a fibre tip collecting a certain solid angle at a wavelength $\lambda$, the numbers of both longitudinal and transverse modes scale as $\lambda^{-2}$~\cite{Spasibko2012}. In total, the number of modes scales as $\lambda^{-4}$, which has been taken into account in the theoretical calculation. One can also notice that the experimental ring-shaped spectrum (panel b) is brighter at large angles. This is caused by the preferable emission along the pump Poynting vector and in the phase matching direction~\cite{giant}.

\begin{figure}[htbp]
\centering
\includegraphics[width=\linewidth]{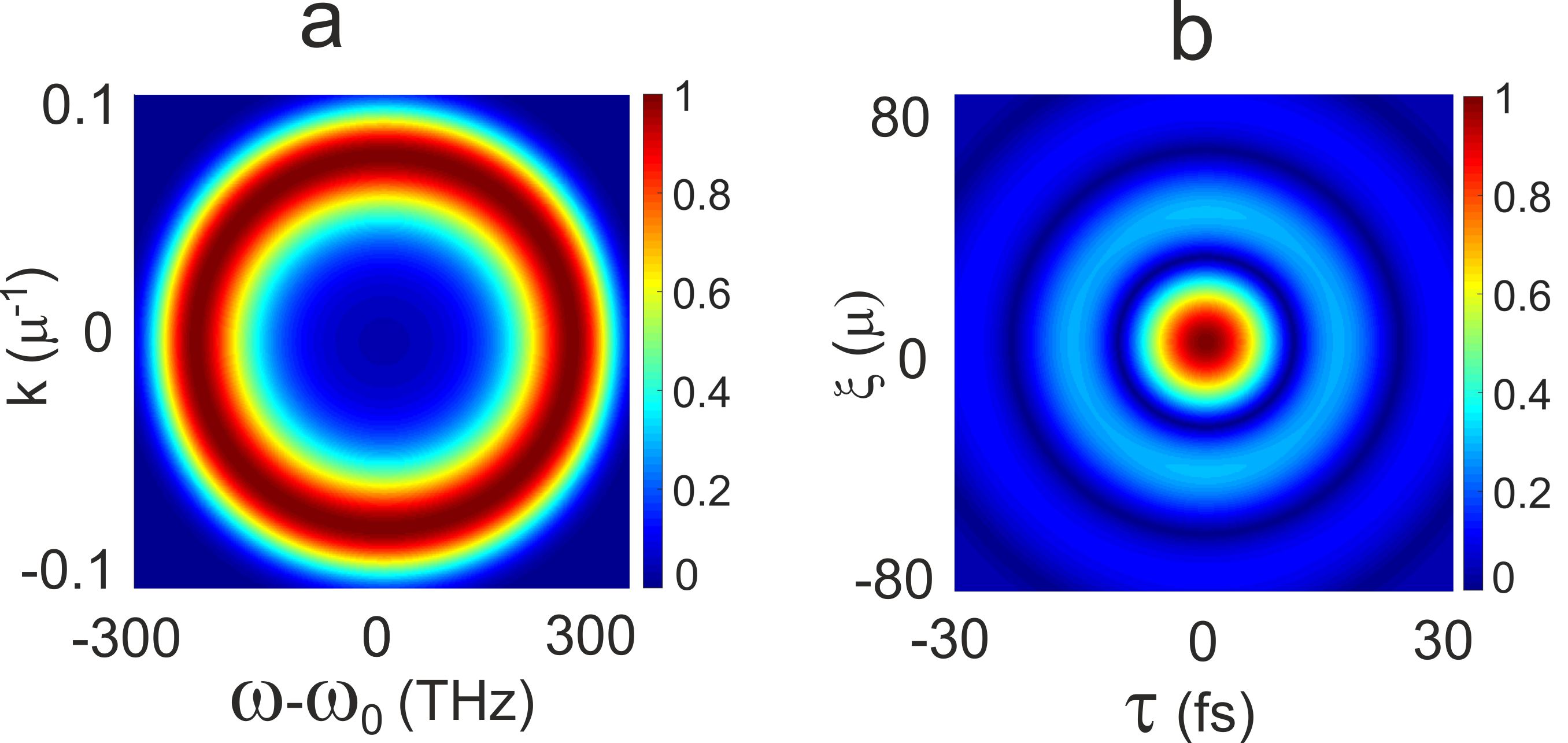}
\caption{Wavelength-angular spectrum as shown in Fig.~\ref{fig:spectra_PDC}d recalculated to coordinates $\omega-\omega_0,k$, where $\omega_0$ is the frequency of exact phase matching (a), and the corresponding first-order correlation function (b).}
\label{fig:spec_g1_high}
\end{figure}

The spectrum as shown in Fig.~\ref{fig:spectra_PDC} b,d acquires an ideal ring-shape if, from the wavelength $\lambda$ and external angle of emission $\theta$, one passes to the cyclic frequency $\omega=2\pi c/\lambda$ and the transverse wavevector $k=\omega\sin\theta/c$. The calculated frequency-wavevector spectrum $S(\omega,k)$ is shown in Fig.~\ref{fig:spec_g1_high}a.

In addition to the spectrum, we have measured the second-order autocorrelation function for collinear frequency-degenerate PDC in the angular (Fig.~\ref{fig:spectra_PDC}e) and wavelength (Fig.~\ref{fig:spectra_PDC}f) domains. In the first measurement, a $100\mu$m slit was scanned in the focal plane of a 200 mm lens, which resulted in an angular resolution of 0.5 mrad. This angular scanning was performed with the radiation filtered by a  bandpass filter with a full width at half maximum (FWHM) of 7 nm around 1600 nm. The obtained distribution shows the peak in the near-collinear direction with a FWHM of 1.3 mrad. The measurement in the wavelength domain was done using a monochromator with the resolution 3.3 nm. The FWHM of the obtained peak is 8.9 nm. In both cases, the theoretical height of the peak is 3, due to the super-bunched radiation statistics~\cite{Iskhakov2012}, but this value was not achieved in experiment due to the insufficient spectral and angular filtering. In context of what follows, it is important that the width of the correlation function is much smaller than that of the spectrum, both in the angle and in the frequency. This indicates a high degree of entanglement, in the low-gain regime given by the ratio of the spectral width and the correlation function width (the Fedorov ratio)~\cite{Fedorov2004}.

To use the photon-number correlations of the PDC radiation, one usually has to filter the spectrum either in frequency or in the angle. This filtering is always accompanied by losses - for instance, if the angular spectrum is filtered by placing an aperture, there still remain uncompensated modes in signal and idler radiation~\cite{two-color}. One can try to filter the eigenmodes of the PDC emission through a projection technique, but it is only for space modes that lossless filtering has been demonstrated~\cite{filtering}. Lossless frequency filtering is much more difficult to realize. If the spectrum is restricted, like in the case we consider, no filtering is necessary.

The fact that the frequency-wavevector spectrum is not factorable in the frequency and wavevector parts, $S(\omega,k)\ne S_f(\omega)S_w(k)$, leads to an unusual type of spatiotemporal coherence~\cite{X-ent1,X-ent2}. In its generalized form, the Wiener-Khinchine theorem relates the spatiotemporal first-order correlation function $G^{(1)}(\tau,\xi)$ to the frequency-wavevector spectrum $S(\omega,k)$ via the two-dimensional Fourier transformation~\,\cite{X-ent1}:
\begin{equation}
G^{(1)}(\tau,\xi)=\iint d\omega dk S(\omega, k)e^{ik\xi-i\omega\tau}.
\label{eq:WKh}
\end{equation}
The spatiotemporal correlation function resulting from a non-factorable frequency-angular spectrum is also not factorable into purely temporal and spatial parts, $G^{(1)}(\tau,\xi)\ne G_t^{(1)}(\tau)G_s^{(1)}(\xi)$. This means that spatial and temporal coherence features are coupled: the reduction of coherence due to a spatial displacement can be compensated by a temporal delay, and vice versa. Such coupling between the spatial and temporal coherence was observed in Refs.~\cite{X-ent1,X-ent2} for the `X-spectrum'. In the case of a ring-shaped spectrum of Fig.~\ref{fig:spec_g1_high}a, the correlation function has a shape shown in Fig. ~\ref{fig:spec_g1_high}b. It has a pronounced central peak corresponding to the trivial case of $\tau=0,\,\xi=0$, but it is surrounded by weaker rings, leading to the non-factorability. In the rings, the lack of temporal coherence and the lack of spatial coherence can compensate for each other.

\begin{figure}[htbp]
\centering
\includegraphics[width=\linewidth]{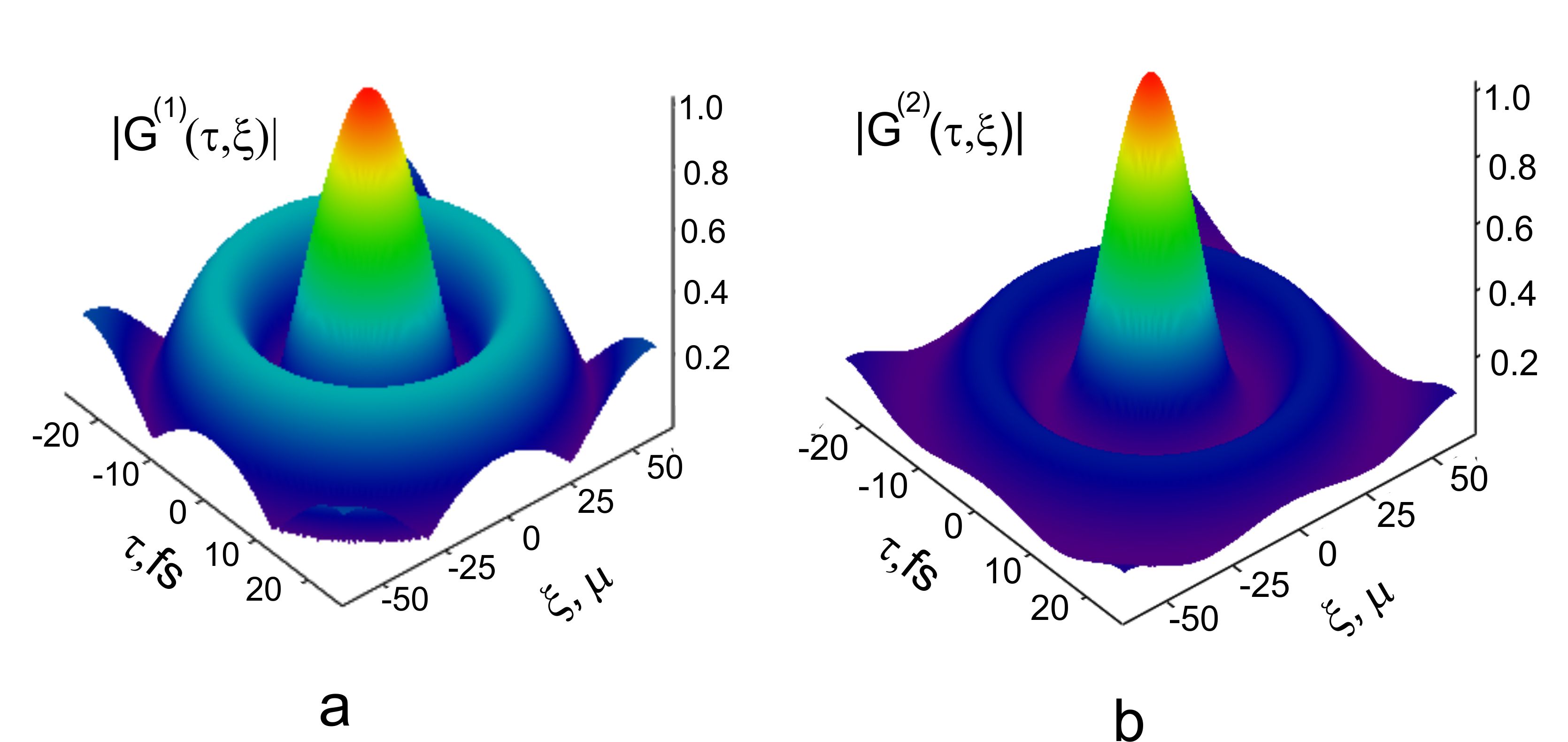}
\caption{The absolute values of the first- (a) and second-order (b) correlation functions calculated for two-photon light with the spectral amplitude given by Eq.~(\ref{eq:ring}).}
\label{fig:G}
\end{figure}

A similar spectrum of PDC will be observed in the low-gain case, where pairs of photons (biphotons) are emitted. The two-dimensional spectrum $S(k,\omega)$ is then determined by the two-photon amplitude (TPA) $F(\omega_s,\omega_i,k_s,k_i)$, which is the probability amplitude for the signal and idler photons to be emitted with the wavevectors $k_s,\,k_i$ and the frequencies $\omega_s,\,\omega_i$, respectively. In the case of the pump pulse long enough and the pump beam waist large enough, the photon pair is highly entangled in wavevector and frequency, hence also in space and time. This corresponds to a high Fedorov ratio, which is indeed the case for the parameters used in our experiment. Then, one can assume $F(\omega_s,\omega_i,k_s,k_i)=F(\omega_s,k_s)\delta(\omega_i+\omega_s-\omega_p)\delta(k_i+k_s)$, with $\omega_p$ being the pump frequency and $F(\omega_s,k_s)$ the spectral amplitude, and~\cite{Chekhova}
\begin{equation}
S(\omega, k)=|F(\omega, k)|^2.
\label{eq:spectrum}
\end{equation}
Using Eqs.~(\ref{eq:WKh},\ref{eq:spectrum}), we have calculated the first-order correlation function with the spectral amplitude taken in the form
\begin{equation}
F(\omega, k)=\frac{\sin(\Delta k L/2)}{\Delta k L/2}e^{i\Delta k L/2},
\label{eq:ring}
\end{equation}
where $L$ is the crystal length and $\Delta k$ the wavevector mismatch. For the parameters used in our experiment, the result is presented in Fig.~\ref{fig:G}a. The first-order correlation function is very similar to the one that should be observed at high gain (Fig.~\ref{fig:spec_g1_high}b).

Even more interestingly, the non-factorability of the frequency and angular (temporal and spatial) degrees of freedom leads to a new type of entanglement, with the spatial degrees of freedom of a photon pair entangled to its temporal degrees of freedom~\cite{X-ent3}. Indeed, for the same case of a highly entangled photon pair, its second-order correlation function, normalized to its maximum value, can be calculated as~\cite{Chekhova}
\begin{equation}
G^{(2)}(\tau,\xi)=|\iint d\omega dk F(\omega, k)e^{ik\xi-i\omega\tau}|^2,
\label{eq:G2}
\end{equation}
where the maximum value of the second-order correlation function is assumed to be so high that the background level (unity in reality) gets close to zero after the normalization. The corresponding shape is shown in Fig.~\ref{fig:G}b.

The figure shows the probability amplitude to have two photons from a pair separated by time $\tau$ and by space $\xi$. Similarly to the case studied in Ref.~\cite{X-ent1}, it has a very pronounced central peak while the ring, showing the coupling between temporal and spatial coherence, is less pronounced. However, in contrast to the first-order correlation function, the second-order correlation function is sensitive to the phase of the spectral amplitude. By imposing on the latter an axially varying phase, one can get rid of the central peak and obtain a photon pair with very unusual properties: the space and time intervals $\xi$ and $\tau$ between the photons of a pair will lie on a circle,
\begin{equation}
\frac{\xi^2}{\xi_c^2}+\frac{\tau^2}{\tau_c^2}=1,
\label{eq:circ}
\end{equation}
where $\xi_c,\tau_c$ are the correlation distance and correlation time, which depend on the phase matching condition. Relation~(\ref{eq:circ}) means that the two photons of a pair are never at the same point at the same time.

The axially varying phase can be imposed on the spectral amplitude, for instance, by using a spatial light modulator (SLM) at the output of a crossed-dispersion scheme~\cite{big} where a two-dimensional distribution as in Fig.~\ref{fig:spectra_PDC}b,d is formed.

\begin{figure}[htbp]
\centering
\includegraphics[width=\linewidth]{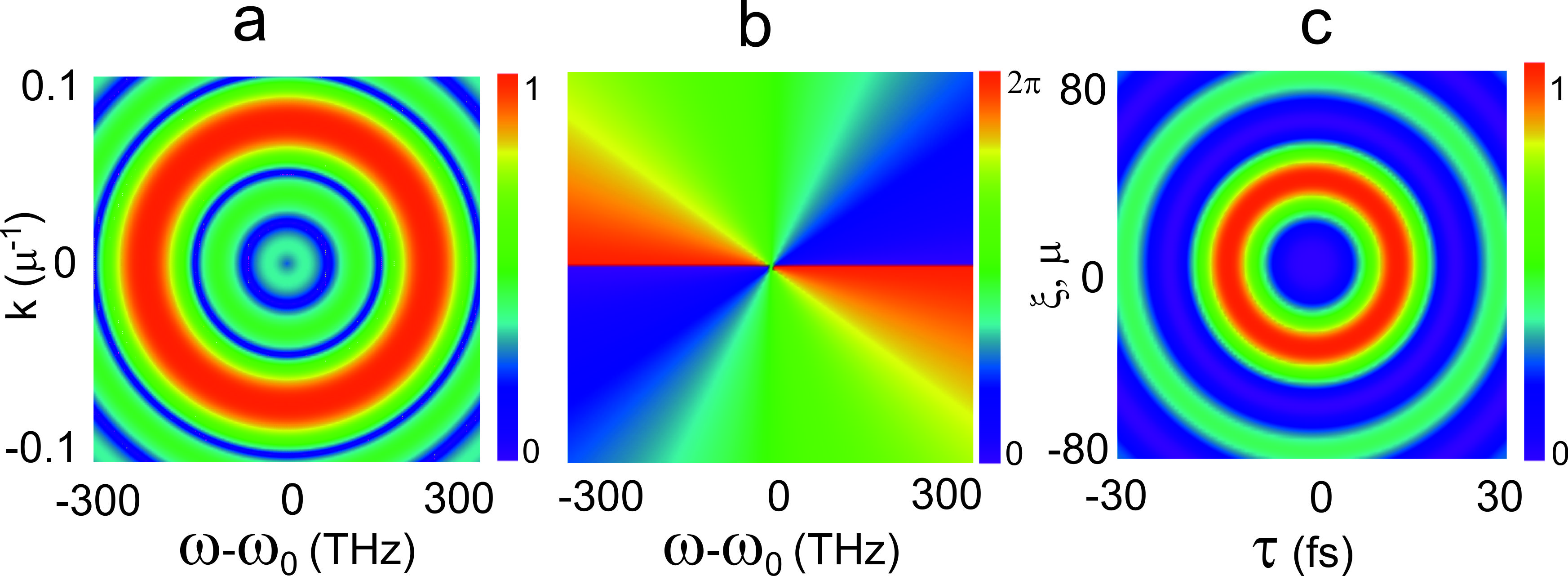}
\caption{The absolute value of the spectral amplitude $F(\omega,k)$ (a), the phase distribution $e^{2i\phi}$ imposed on it (b), and the absolute value of the resulting correlation function $G^{(2)}(\tau,\xi)$ (c).}
\label{fig:phase}
\end{figure}

Fig.~\ref{fig:phase} shows an example of such a correlation function. Panel a presents the absolute value of a ring-shaped spectral amplitude as a function of the frequency detuning $\omega-\omega_0$ from exact degeneracy and the transverse wavevector $k$, calculated for low-gain PDC with the crystal oriented the same way as in Fig.~\ref{fig:spectra_PDC}c,d. Panel b shows the phase distribution over the frequency and transverse wavevector, which can be imposed with the help of an SLM. Namely, the spectral amplitude $F(\omega,k)$ was replaced by
\begin{equation}
F'(\omega,k)=F(\omega,k)e^{2i\phi},
\label{eq:phase}
\end{equation}
where $\phi$ is the azimuthal angle in panel a. (In general, any phase modulation $e^{in\phi}$ will lead to a similar result; here we chose $n=2$ just as an example.) Finally, panel c shows the resulting correlation function, which indeed has the shape of a ring given by Eq.~(\ref{eq:circ}), with $\tau_c=13$ fs, $\xi_c=35\,\,\mu$m. Note that the size of the ring can be increased if the azimuthal phase distribution has faster oscillations, $F'(\omega,k)=F(\omega,k)e^{in\phi}$ with integer $n>2$, but for any $n\ne0$ the central peak in the second-order correlation function will disappear. One can notice that the same phase modulation will also eliminate the central peak in the 'X-type' correlation function of Ref.~\cite{X-ent3}.

This unusual type of correlations, with the two photons forming pairs but always separated either in space or in time, or both, can be useful if one needs to avoid two-photon effects like absorption. On the other hand, the original single-peaked correlation function, like in Fig.~\ref{fig:G}b, can be restored by the appropriate phase modulation if necessary. This fact can be used in time-and space-resolved nonlinear spectroscopy with two-photon light or bright squeezed vacuum. For instance, by varying the size of the ring in Fig.~\ref{fig:phase}c and observing a two-photon effect (absorption, second-harmonic generation, ionization, etc.) one can measure the nonlinear response time of the material.

In conclusion, we have experimentally demonstrated PDC with a bounded broadband frequency-angular spectrum, which can be turned into a ring-like spectrum. This type of spectrum, first of all, provides a possibility to collect all radiation without losses, which is an advantage in experiments on squeezed light detection. In addition, a ring-shaped spectrum leads to an interesting type of spatio-temporal coherence, with the temporal coherence coupled to the spatial one. Furthermore, for two-photon light emitted under such phase matching, the second-order correlation function is of unusual type, manifesting entanglement between temporal and spatial degrees of freedom. With the help of a phase shift imposed on the two-photon amplitude, the second-order correlation function can acquire a ring shape, describing two photons that are never at the same place at the same time. This fact can have important applications in time-and space-resolved nonlinear spectroscopy.

\section*{Funding Information}
This work was funded by the Seventh Framework Program (FP7) (308803, project BRISQ2).

\section*{Acknowledgments}
We would like to dedicate this paper to the memory of A. N. Penin who observed the 'X-spectrum' in the 1970-s and has been inspiring many of our experiments until very recently.
% Bibliography
%\bibliography{sample}

\end{document}